# Control of resonant ionization as a function of time delay between two XUV few-femtosecond pulses. Quantitative application to helium

Theodoros Mercouris[1], Yannis Komninos[2] and Cleanthes A. Nicolaides[3]

*Theoretical and Physical Chemistry Institute, National Hellenic Research Foundation, 48 Vasileos Constantinou Avenue, Athens 11635, Greece*

1) thmerc@eie.gr  2) ykomn@eie.gr  3) caan@eie.gr

**August 20, 2019**

**Abstract:** By obtaining and using nonperturbative solutions of the time-dependent Schrödinger equation for a time-dependent *interference* scheme involving the process of two-photon resonant ionization of Helium via the $1s2p\ ^1P^o$ (58.4 nm) and $1s4p\ ^1P^o$ (52.22 nm) excited states, we show that it is possible to use few-femtosecond positive or negative *time delays*, $\Delta t$, between two XUV Gaussian few-femtosecond pulses of moderate intensities in order to control the probability of ionization at clearly defined exit energies corresponding to the sums of photon frequencies ($\omega_1 + \omega_1$), ($\omega_2 + \omega_2$), and ($\omega_1 + \omega_2$, or $\omega_2 + \omega_1$).

The calculations used wavefunctions which are *state-specific* for the discrete spectrum, (up to the $1s7g\ ^1G$ Rydberg state), as well as for the energy-normalized continuous spectrum, (up to 2.0 a.u. above threshold with angular momenta $\ell = 0, 1, 2, 3, 4$). For this system, using $\Delta t$ as a control knob, the effects of interference on the photoelectron spectrum are determined clearly for pulses with field-cycles ranging from about 15 to about 80 cycles. The analysis has included the comparison of the nonperturbative results, obtained by implementing the *state-specific expansion approach*, with those obtained from the application of second order time-dependent perturbation theory with two Gaussian pulses of finite duration.

# I. TIME DELAY AS A CONTROL PARAMETER IN A TWO-PHOTON RESONANT IONIZATION PROCESS WITH TWO XUV ULTRASHORT PULSES

Current experimental/technological work in many facilities by large teams around the world aims at improving the performance and versatility of sources of radiation pulses that can produce ultrashort pulses at different ranges of wavelengths, having weak, moderate or strong intensities, say between $10^{11}$ and $10^{14}$ W/cm$^2$. This work involves the production and application of either table-top, high-harmonic generation (HHG)-based pulses, or, of well-characterized pulses from free-electron laser (FEL), whose duration is in the range of a few decades of femtosecond (*fs*) down to decades of attosecond (*as*). In principle, such pulses are suitable for use in a variety of new types of spectroscopic studies in atomic, molecular and optical (AMO) physics.

These developments, in conjunction with corresponding advances of theory and of many-electron methods that can deal quantitatively with the solution of the many-electron time-dependent Schrödinger equation (METDSE) and with its appropriate utilization, have ushered AMO physics into a new era of spectroscopy, where the physically relevant information about electron dynamics with respect to ultrashort changes of time can be understood not only phenomenologically or descriptively, but even quantitatively for real systems.

Brief discussions on aspects of the progress that has thus far been achieved experimentally and on prospects for applications to various problems and areas of physics and chemistry using such pulses and corresponding spectroscopic techniques, were recently presented in a compendium of short articles by many authors in [1]. The generation and application of well-characterized FEL pulses of short wavelengths, say in the extreme ultraviolet (XUV) and beyond, is expected to open new vistas in the study of time-resolved electron dynamics.

Of recent interest to us [2] was the quantitative study and analysis of the *absolute cross-sections* of the two-XUV photon ionization of Helium, on- and off-resonance with the Rydberg states $1s2p\ ^1P^o$, $1s3p\ ^1P^o$ and $1s4p\ ^1P^o$, which were obtained by averaging the time-dependent probabilities that emerged from the nonperturbative solution of the METDSE. The impetus for that study came from the publication of the first measurements of these quantities using XUV pulses from the FEL in RIKEN, Japan [3, 4]. The results of [2] were compared with the experimental values and with earlier theoretical results from perturbative as well as nonperturbative time-independent calculations. For the conclusions the reader is referred to [2].

The experimental work of Fushitani et al [4] also included the investigation of aspects of the spectroscopy which is possible when two-color *fs* pulses, with time delay $\Delta t$, are used for the study of the ionization of Helium. They combined an ultrashort optical laser pulse of 268 *nm* (4.63 eV) with FEL pulses of 59.7 *nm* (20.8 eV). The FEL pulse was off-resonance from the $1s2p\ ^1P^o$ state by 0.4 eV.

In view of the developments discussed in [1-4] and in their references, a question which is relevant to time-resolved electron dynamics is the following: What kind of new information can be extracted when, instead of one, two ultrashort pulses with XUV or shorter wavelengths are generated and used in time-dependent spectroscopy with atoms or with their positive ions? For example, in a recent publication co-authored by 40 scientists, Prince et al [5] reported experimental results on the *coherent control* [6] of an ionization process in Neon at the level of only a few *as* time-resolution, using two XUV ( 63.0 *nm* and 31.5 *nm*) *fs* pulses and adjusting their *phase difference*, $\varphi$, so as to control the asymmetry of the photoelectron angular distribution. The authors of [5] argued that their experimental demonstration '*opens the door to new short-wavelength coherent control experiments with ultrahigh time resolution and chemical sensitivity*', (abstract of [5]), with prospects of new types of application, '*in analogy to what happened in the last few decades in the field of optical laser-based research*'. (Prince and Masciovecchio, on page 16 of [1]).

The theoretical work reported here has taken our earlier investigations [2] to a more complex and challenging level, by exploring quantitatively the possibility of *control* of a two-photon resonant ionization process in Helium, when two different XUV pulses with ultrashort duration are used. As before with optical laser-based research on processes of photodissociation and of photoionization, the underlying principle is the presence of *interference* of different excitation paths ending at the same final state in the continuum [6]. However, the solution of the problem requires going beyond the phenomenology of the mechanism of interference, and obtaining and using time-dependent solutions of the METDSE that account for the interplay between electronic structures, state-mixings, and time-dependent dynamics. This endeavor is analogous to the one used in previous investigations having to do with AC fields, where, however, the many-electron perturbative or nonperturbative treatments can be done in terms of time-independent formulations. For example, see [7] for theory and ab initio results obtained from many-electron calculations on the functional dependence of the '*interference generalized cross-section*' on the *phase differences* between dichromatic or trichromatic weak AC fields. We note that the use of the *relative phase* as a control parameter in two-color studies of spectroscopy where path interference plays a crucial role, continues to find new areas of application, e.g., [8].

In the present problem, interference is achieved by irradiating Helium not by one but by two XUV ultrashort Gaussian pulses of moderate intensities, $1 \times 10^{12}$ W/cm$^2$ and $8 \times 10^{12}$ W/cm$^2$, which are applied with *time delay*, $\Delta t$, and whose wavelengths are chosen so as to cause two-photon resonant ionization through the transitions $1s2 \, {}^1S_0 \rightarrow 1s2p \, {}^1P_1^o$ at 58.4 *nm* and $1s2 \, {}^1S_0 \rightarrow 1s4p \, {}^1P_1^o$ at 52.22 *nm*. Obviously, the two resonant paths towards the same final state of the continuum, depicted in lowest order as,

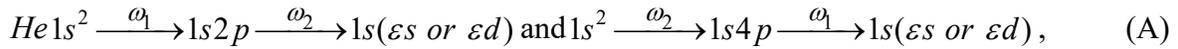

$$He\,1s^2 \xrightarrow{\;\omega_1\;} 1s2p \xrightarrow{\;\omega_2\;} 1s(\varepsilon s \text{ or } \varepsilon d) \text{ and } 1s^2 \xrightarrow{\;\omega_2\;} 1s4p \xrightarrow{\;\omega_1\;} 1s(\varepsilon s \text{ or } \varepsilon d), \qquad (A)$$

can interfere. ($\omega_1 = 0.78$ a.u. and $\omega_2 = 0.87$ a.u. are the central frequencies of the two pulses). The question we asked is whether, and under what conditions, such an interference can be induced with observable results by varying $\Delta t$.

We note that the two-photon resonant interference scheme was first proposed by Chen, Shapiro and Brumer [9, 6] in the era of optical lasers and AC fields, and was first demonstrated experimentally by Pratt [10], followed by Wang and Elliott [11]. Their analysis was based on the time-independent formula of lowest order perturbation theory and on consideration of quantities such as the degree of detuning, laser power, or relative polarization, as control parameters [10,11].

In the present case, the theoretical treatment must necessarily involve time explicitly. The relevant information is obtained from systematic calculations that solve the METDSE nonperturbatively, using a two-color perturbation $V_{\text{int}}(t) = V_1(\omega_1, t) + V_2(\omega_2, t + \Delta t)$. Now, it is $\Delta t$ that has the role of the control knob.

## II. CONSTRUCTION OF THE PROBLEM AND QUANTITATIVE SOLUTION

The proposal of the two-photon resonant ionization of Helium using two ultrashort XUV pulses which was outlined in the Introduction, is displayed in Fig. 1. The main goal was to identify quantitatively, from first principles, a set of parameters of two XUV Gaussian pulses and the corresponding values of ultrashort time delays, $\Delta t$, that can be used to affect in a significant and controllable way the photoelectron spectrum via the effect of interference. As discussed in section III, this has proven possible. The duration of either pulse is in the range of a few femtoseconds, for intensities in the range $1 \times 10^{12}$ - $1 \times 10^{13}$ W/cm$^2$. Calculations also showed that, for intensities close to $10^{14}$ W/cm$^2$, the resulting spectrum loses its clarity, indicating the presence of strong shifting and mixing of states from both the discrete and the continuous spectrum.

The photoelectron peak around the energy $E_{12} = \omega_{1} + \omega_{2}$ - $E_{ion}$, is essentially independent of the phase difference, $\varphi$, because the paths that interfere at $E_{12}$ have the same dependence on it. This is a convenient condition for experimentalists, because the steady control of $\varphi$ for FEL pulses of high energy is not achievable easily.

On the other hand, we expect that $\Delta t$ should play a significant role in the vicinity of $E_{12}$, because, for ultrashort pulses, the population transfer to the intermediate states $1s2p\,^1P^o$ and $1s4p\,^1P^o$ is strongly dependent on it. It follows that the magnitudes of the photoelectron peaks corresponding to the photon frequencies $(\omega_1 + \omega_1)$, $(\omega_2 + \omega_2)$, and $(\omega_1 + \omega_2$, or $\omega_2 + \omega_1)$, will also depend on $\Delta t$ and, of course, on intensity, in a way which cannot be predicted without accurate calculations that solve the METDSE with a two-color time-dependent interaction operator, $V_{int}(t)$.

The basic step towards the quantitative understanding of the electron dynamics associated with the processes of Fig. 1 is the calculation of the solution of the METDSE from first principles. This was done nonperturbatively by implementing the *state-specific expansion approach* (SSEA) [12]. For reasons of comparison and for additional insight into the role of the continuum and of discrete states other than the $1s2p\,^1P^o$ and the $1s4p\,^1P^o$, the METDSE was also solved at the level of second order time-dependent perturbation theory (SOTDPT) with Gaussian pulses of finite duration – see below.

The fundamental equations are,

$$\boldsymbol{H}(t)\Psi(t) = i\hbar\frac{\partial\Psi(t)}{\partial t} \tag{1a}$$

$$\boldsymbol{H}(t) = \mathbf{H}_{atom} + \boldsymbol{V}_{int}(t), \qquad \boldsymbol{V}_{int}(t) = \boldsymbol{V}_1(\omega_1, t) + \boldsymbol{V}_2(\omega_2, t + \Delta t) \tag{1b}$$

In our work, the coupling operator in $\boldsymbol{V}_{int}(t)$ is the full multipolar electric Hamiltonian, $H_{el}$,

$$H_{el} = e\sum_j\int_0^1\vec{r}_j\cdot\vec{E}(\lambda\vec{\kappa}\cdot\vec{r}_j)d\lambda \tag{2}$$

The compact form (2) is taken from Eq. (5.35) of Loudon's book [13]. The reasons for choosing (2) and the theory which deals with the properties of this operator when calculating *bound-bound*, *bound-free* and *free-free* matrix elements in spherical symmetry, can be found in [14] and its references. As before, the required matrix elements are computed by applying the electric dipole selection rules for linearly polarized light.

In accordance with Fig. 1, the $He\,1s^{2\,1}S$ state is assumed to interact with two XUV time-delayed pulses. The pulse forms are,

$$E(t) = \sum_k E_k(t), \qquad E_k(t) = F_k e^{-\alpha_k(t-t_k)^2}\cos(\omega_k t), k = 1,2 \tag{3}$$

$$\alpha_k = \frac{2\ln(2)}{\tau_k^2}, \qquad \tau = \text{full width at half-maximum} \tag{3a}$$

The time delay is positive or negative, and is defined as,

$$\Delta t = t_2 - t_1 \tag{4}$$

The final results (section III) determine the photoelectron probability distribution, (per atomic unit of energy),

$$\frac{dP_\varepsilon}{d\varepsilon} = \sum_\ell \left| \alpha_{\varepsilon,\ell}(T_{end}) \right|^2 \tag{5}$$

where $\alpha_{\varepsilon,\ell}$ is the coefficient of the energy-normalized state of the continuum at energy $\varepsilon$, with angular momentum $\ell$, and $T_{end}$ is the time at which the interaction is essentially zero.

The bulk of the results reported in the paper were obtained for field intensities $I_k$ of the order of $10^{12}$ W/cm$^2$. They are clear and demonstrate the physics quantitatively. In particular, the values of $F_k$ ($I_k$) for which the results presented in Figs. 2-5 were obtained are,

$$F_1 = 0.00534 \, a.u. \, (I_1 = 1 \times 10^{12} \, W/cm^2) \text{ and, } F_2 = 0.015 \, a.u. \, (I_2 = 8 \times 10^{12} \, W/cm^2) \tag{6}$$

The intensity for the path through the state $1s4p \, ^1P^o$ (52.2 $nm$), was chosen to be larger because the corresponding dipole matrix element is smaller than the one for $1s^2 \, ^1S_0 \rightarrow 1s2p \, ^1P_1^o$ (58.4 $nm$).

We also carried out calculations for higher intensities, for which, as expected, the field-induced shifts and higher-order state-mixings are enhanced and cause a blurring of the photoelectron spectrum. In this way, we are able to establish a set of values for intensities and for pulse durations for which possible future work on this system could test with confidence the theoretical predictions as well as the experimental techniques.

## A.    Calculations

The $\Psi(t)$ of Eq.(1) was calculated at two different levels of theory. In both cases, the relevant *state-specific* electronic structures, electron correlations, and energy-normalized scattering states were taken into account systematically and reliably. The characteristics of the pulses were the same for both theoretical models.

The first level, which is the generally applicable one for problems of time-resolved electron dynamics for strong and/or ultrashort pulses, has to do with the nonperturbative calculation of $\Psi(t)$. The SSEA results from this type of solution serve as a reliable reference for both experiment and theory.

The second level is that of SOTDPT, where the Gaussian forms of the two pulses and their finite duration of a few femtoseconds were taken into account explicitly.

For the SSEA calculation, the time-dependent wavefunction, $\Psi_{SSEA}(t)$, was constructed and computed in the form,

$$\Psi_{SSEA}(t) = \sum_{\ell=0,\, n=\ell+1} \alpha_{n,\ell}(t)\Phi_n(1sn\ell\,^1L) + \sum_{\ell=o}\int d\varepsilon\,\alpha_{\varepsilon,\ell}(t)\Phi_\varepsilon(1s\varepsilon\ell\,^1L) \qquad (7)$$

where $\Phi_n$ and $\Phi_\varepsilon$ are the symmetry-adapted 2-electron bound and energy-normalized scattering wavefunctions for the *He* states, labeled by each reference configuration.

As explained in [12] and its references, the choice of the N-electron wavefunctions in the SSEA is specific to the electronic structure of each state in the expansion, (discrete, resonance and purely scattering state), and to each problem. The theoretical formulations and methods for their calculation at different levels of approximation, depend on the property/phenomenon of interest. In general, these wavefunctions are constructed from separately optimized *numerical* and *analytic* one-electron functions, [15] and its references.

In the present case of Helium, the wavefunctionsin expansion (7) were, in large part, already calculated and used in the recent work of [2]. In order to eliminate any numerical inaccuracies and uncertainties, especially when it comes to the diffuse Rydberg wavefunctions, the state-specific bound wavefunctions were computed and used (in matrix elements) *numerically*, by solving, for each discrete state $\Phi_n$, the multi-configurational Hartree-Fock (MCHF) or HF (for the Rydberg states) equations, using the code published by Froese-Fischer [16]. The energy-normalized scattering orbitals were obtained numerically in the frozen core of the $He^+$ 1s state.

The ground-state wavefunction, $\Phi(1s^2\,^1S)$, was obtained from a numerical MCHF calculation with the $1s^2$, $2s^2$, $2p^2$, $3s^2$, $3p^2$, $3d^2$, $4s^2$, $4p^2$, $4d^2$, $4f^2$ configurations, essentially exhausting the relevant to the dynamics contributions of electron correlation. The excited discrete states are of the Rydberg type, $1sn\ell\,^1L$. They were represented by state-specific numerical HF orbitals with $n = 2, 3, 4, 5, 6, 7$ and $\ell = 0, 1, 2, 3, 4$, i.e., we included all discrete states up to $1s7g\ {}^1G$.

Using energies of the energy-normalized scattering orbitals $\varepsilon\ell$ in the range from 0 to 2.0 a.u. above threshold, and angular momenta with $\ell = 0, 1, 2, 3, 4$, the convergence of the SSEA calculations was very good. The number of coupled equations corresponding to the converged expansion (7) was of the order of 10000, the overwhelming majority involving, as usual, matrix elements with states from the continuum.

## III.  RESULTS FROM THE SSEA AND FROM THE TIME-DEPENDENT PERTURBATION THEORY TO SECOND ORDER

Following a series of exploratory SSEA calculations, we identified sets of pulse parameters that can demonstrate with clarity and accuracy the physics of the problem. Specifically, we delineated roughly the region of moderate intensities that produce photoelectron spectra, (probability distribution per atomic unit of energy), where every peak is interpreted with certainty, (Figs. 2-5), from the region of strong fields, where the spectrum is altered by being broadened while acquiring additional smaller peaks (Fig. 6).

For the moderate intensities of (6), we will discuss three sets of results, presented as Figs. 2-5. These sets correspond to pulses of about 20, 40 and 80 field-cycles.  For the strong field strengths of $I_1 = 1 \times 10^{14}$ W/cm$^2$ and $I_2 = 2 \times 10^{14}$ W/cm$^2$, the results are reported in our Fig. 6. For the reasons given below, our study also included complementary calculations from the application of SOTDPT.

### A.  Pulses of about 20 field-cycles

The results of the first set with the moderate intensities (6) are displayed in Fig. 2. They were obtained for $\tau_k$, $k = 1, 2$, of the order of 20 field cycles (duration of about $\tau_1 = 4\,fs$ and $\tau_2 = 3\,fs$). The left and right peaks correspond to the absorption of two $\omega_1$ and two $\omega_2$ photons respectively. These peaks are essentially independent of $\Delta t$.

However, the absorption of one photon from each pulse gives rise, through the interference of different paths, (Fig. 1), to the peak in the middle, whose dependence on $\Delta t$ is found to be significant: For negative $\Delta t$, of the order of 2-5 $fs$, where the $\omega_2$ pulse precedes the $\omega_1$ pulse, the height of this peak gradually decreases with respect to its value for $\Delta t = 0$. On the contrary, for positive $\Delta t$, the height of the peak increases.

This observable dynamical effect, which was obtained quantitatively from first principles, constitutes the main prediction of this work. Its explanation is as follows:

The choice of the pulse intensities, ($F_2 > F_1$), is such that the Rabi frequencies for both resonant couplings ($1s^2 \xrightarrow{\omega_1} 1s2p$, $1s^2 \xrightarrow{\omega_2} 1s4p$) are essentially the same, which means that the population transfer from $1s^2$ to the two $^1P^o$ states is nearly the same. In the case of (relatively) large negative time delays, i.e., when the pulses do not overlap, the $\omega_1$ pulse starts to act following the fading out of the interaction due to the $\omega_2$ pulse. Hence, the transfer of population from the $1s4p\ ^1P^o$ state to the continuum states $1s\varepsilon s\ ^1S$ and $1s\varepsilon d\ ^1D$ is much weaker than the one resulting when the opposite case holds, namely, when the time delays become positive and large, in which case it is the $\omega_2$ pulse that pumps population from $1s2p\ ^1P^o$ to the continuum. This is so because the dipole transition matrix elements $<1s2p\ ^1P^o \mid \mathbf{H}_{el} \mid 1s\varepsilon s,\ 1s\varepsilon d >$ are larger than the $<1s4p\ ^1P^o \mid \mathbf{H}_{el} \mid 1s\varepsilon s,\ 1s\varepsilon d >$ ones, and, in addition, the field strengths satisfy $F_2 > F_1$. Therefore, for time delays in the range between (relatively) large negative and large positive values, or, equivalently, in the range between non-overlapping pulses, the height of the interference (middle) peak varies continuously.

In addition to the understanding of the phenomenon that is presented in Fig. 2 in terms of matrix elements and field strengths, it is also of interest to obtain additional quantitative information on how the phenomenon is affected by the states of the full spectrum as a function of the duration (number of field cycles) of the two pulses. We recall that the spectral widths of these pulses are not negligible. For example, for the present case of 20 cycles, these spectral widths are 0.76 eV and 0.62 eV, (depicted on Fig. 1), when the energy differences between the two $^1P^o$ excited states of interest from their neighboring $^1P^o$ states are, $\Delta E(1s2p, 1s3p) = 1.87$ eV, and $\Delta E(1s4p, 1s3p) = 0.65$ eV, $\Delta E(1s4p, 1s5p) = 0.30$ eV.

The information concerning the understanding of the participation of the states of the spectrum to the proper description of the interference peak can be gained, in principle, by running a series of SSEA calculations with different expansions, since the structure of the theory allows in a straight forward way to assess the degree of importance of each state-specific wavefunction. However, this is a computationally very time consuming enterprise. Therefore, we turned to much more economic calculations at the level of SOTDPT, i.e., at the level implied by the scheme (A), where only a subset of the states of the full spectrum is involved. These results were then compared to the SSEA nonperturbative results which are presented in Figs. 2-6. This comparison sheds light to the understanding of the significance of the contribution to the formation of the interference peak of those states that are not on exact resonance with the center of the two pulses, and of the range of pulse durations and field intensities for which the present problem can be tackled reliably in terms of the SOTDPT.

The formula for the matrix element of the SOTDPT can be found in textbooks, e.g., in Chapter 2 of [17]. In the interaction picture, it is formally written as the operator,

$$A^{(2)}(t) = (i/\hbar)^2 \int_{-\infty}^{t} dt_1 V_{\text{int}}(t_1) \int_{-\infty}^{t_1} dt_2 V_{\text{int}}(t_2)$$

Its implementation to the present case of the *He* interference peak reads,

$$\left\langle 1sE_{12}\ell\, ^1L^o \left| A^{(2)}(t) \right| \Phi(1s^2) \right\rangle =$$

$$\left(\frac{-i}{\hbar}\right)^2 \sum_J \int_{-\infty}^{t} dt_1 e^{i(E_{12}-\varepsilon_J)t_1} \left\langle 1sE_{12}\ell\, ^1L^o \left| E(t_1)z \right| 1sJp\, ^1P^o \right\rangle \times \int_{-\infty}^{t_1} dt_2 e^{i(\varepsilon_J-\varepsilon_{1s^2})t_2} \left\langle 1sJp\, ^1P^o \left| E(t_2)z \right| \Phi(1s^2) \right\rangle$$

$$= \left(\frac{-i}{\hbar}\right)^2 \sum_J \left\langle 1sE_{12}\ell\, ^1L^o \left| z \right| 1sJp\, ^1P^o \right\rangle \times \left\langle 1sJp\, ^1P^o \left| z \right| \Phi(1s^2) \right\rangle \times \int_{-\infty}^{t} dt_1 e^{i(E_{12}-\varepsilon_J)t_1} E(t_1) \times \int_{-\infty}^{t_1} dt_2 e^{i(\varepsilon_J-\varepsilon_{1s^2})t_2} E(t_2)$$

$$(8)$$

where $J$ spans the discrete and the continuous parts of the $^1P^o$ spectrum, and $\ell = s, d$.

For reasons of economy, we present only the case with $\Delta t = 0$. By comparing the results of the SOTDPT and of the SSEA, the following conclusion is reached: The height of the interference peak from the SSEA calculation is $2.5 \times 10^{-4}$. If we consider the case where $J$ spans the states of the $^1P^o$ discrete spectrum only, then the SOTDPT result is $1.7 \times 10^{-4}$. However, when the continuous spectrum is included, the result is essentially the same as that of the SSEA. Hence, we have a quantitative understanding of the significance of the scattering states for these field parameters.

### B. Pulses of about 40 field-cycles

The results of the second set are shown in Figs. 3 and 4. They were obtained for $\tau_1 = 8\,fs$ and $\tau_2 = 7\,fs$, durations that correspond to about 40 field cycles. The analysis given for the results of Fig. 2 is also valid for this case. Now, the spectral widths are smaller. They are 0.38 eV and 0.31 eV respectively. Therefore, one should expect that the contribution to the two-photon resonant ionization process of the $1snp \ ^1P^o$ discrete states with $n \neq 2,4$, as well as of the $1s\varepsilon p \ ^1P^o \ \varepsilon \geq 0$ scattering states, should be less important. Indeed, for $\Delta t = 0$, the SSEA value of the height of the interference (middle) peak is $2.75 \times 10^{-3}$, while the value from Eq. (8) with $J$ spanning only the discrete $^1P^o$ spectrum is $2.63 \times 10^{-3}$. The contribution of the continuous part of the $^1P^o$ spectrum is smaller than before. The total value of the SOTDPT calculation is almost the same as that from the SSEA calculation.

### C. Pulses of about 80 field-cycles

The results of the third set are shown in Fig. 5. They were obtained for $\tau_k$, $k = 1,2$, of about 80 field cycles (duration of about $\tau_1 = 16\,fs$, $\tau_2 = 15\,fs$), for $\Delta t = 0$. Now, the spectral widths are even smaller, with values 0.19 eV and 0.15 eV respectively. Yet, a rather significant difference between the SSEA and the SOTDPT results was obtained. In the former case (SSEA), the height of the interference peak is $3.5 \times 10^{-2}$, whilst in the latter case (SOTDPT), with $J$ spanning the full $^1P^o$ spectrum, the height is $4.2 \times 10^{-2}$. This means that, for the system of interest, the SOTDPT starts to fail for pulses with $\tau$ of about 80 field cycles.

This fact is rationalized as follows: The formalism for AC fields corresponding to Eq.

(8), (i.e., $E(t) = \sum_{k=1}^{2} E_k(t) = \sum_{k=1}^{2} F_k \cos(\omega_k t)$, and $E(-\infty) = 0$, - adiabatic switch-on of the fields), shows that the transition amplitude to the interference (middle) peak is equal to,

$$\left\langle 1sE_{12}\ell \,^1L^o \left| A^{(2)}(\infty) \right| \Phi(1s^2) \right\rangle = 2\pi \left( \frac{-i}{\hbar} \right)^2 \left( \frac{F_1}{2} \right) \left( \frac{F_2}{2} \right) \delta(\varepsilon_{1s^2} - E_{12} + \omega_1 + \omega_2) \times$$

$$\sum_{k=1,J}^{k=2} \frac{\left\langle 1sE_{12}\ell \,^1L^o \left| z \right| 1sJp \,^1P^o \right\rangle \left\langle 1sJp \,^1P^o \left| z \right| \Phi(1s^2) \right\rangle}{\varepsilon_{1s^2} + \omega_k - \varepsilon_J}$$

(9)

This amplitude diverges when resonant coupling occurs. Therefore, the discrepancy between the nonperturbative SSEA and the SOTDPT results indicates that for pulses with $\tau_k$, $k = 1, 2$, of about 80 field cycles or more, the pulses start approaching rapidly the AC limit. In this limit, the time-dependent perturbation theory to lowest order in inadequate in describing quantitatively, even for moderate intensities, the interference phenomenon caused by the two- XUV photon resonant ionization process discussed here.

### D. Pulses of about 40 field-cycles and strong fields

Finally, we turn to a case where the fields of the two pulses are rather strong. Fig. 6 displays the results for $I_1 = 1 \times 10^{14} \, W/cm^2$ and $I_2 = 2 \times 10^{14} \, W/cm^2$, for $\tau_1 = 8 \, fs$, $\tau_2 = 7 \, fs$, (about 40 cycles). Again, $\Delta t = 0$.

With respect to the requirement of securing good convergence, the SSEA calculation takes a considerable amount of time on a medium-size computer. Yet, we thought that it was necessary, in order to explore the limits where the photoelectron spectrum for this problem starts losing the easily interpretable picture of Figs. 2-5. Indeed, although the spectrum of Fig. 6 is reminiscent of the basic features of the three peaks of Figs. 2-5, the broadening, the asymmetry and the additional smaller peaks render its straight forward interpretation precarious and without direct utility for the problem of control which we have studied here.

An additional useful piece of information from the calculation with such pulse parameters is the possibility of testing the validity of the SOTDPT. The results show that, for such intensities, the phenomenon of the interference peak cannot be described reliably by the SOTDPT. Indeed, the SOTDPT value for the top of the peak is 6.6, deviating by a factor of about 60 from the result obtained from the nonperturbative SSEA.

In view of the results and discussion of the previous subsections, we close this section with a brief commentary on the Rabi frequency and its possible effect on the results:

The results that are displayed as Figures 2-5, demonstrate the proposed phenomenon of interference as a function of the (positive or negative) *time delay*, $\Delta t$, with clarity. In view of the physics of Rabi oscillation, it is useful to see how this is achieved. The field strengths were chosen to be $F_1$= 0.00534 a.u. and $F_2$= 0.015 a.u., so as to obtain Rabi frequencies that are almost equal, $\Omega_1$= $F_1$<1s²|z|1s2p>= 0.0022 a.u. and $\Omega_2$= $F_2$<1s²|z|1s4p>= 0.002 a.u.. These frequencies correspond to a Rabi cycle of about 70 *fs*.

We are now in a position to estimate the number of Rabi oscillations, $N_{os}$, for pulses that have Gaussian temporal shape, by using the formula first presented in [18],

$$N_{os} = F_i V_i \frac{1}{2\sqrt{2\pi \ln(2)}} t_i, \qquad i = 1,2 \qquad (10)$$

where, $V_1 = \left\langle 1s^2 \mid z \mid 1s2p \right\rangle$ *and* $V_2 = \left\langle 1s^2 \mid z \mid 1s4p \right\rangle$ .

For all cases shown in Figs. 2-5, the number of Rabi oscillations is much smaller than 1. This fact explains our choice of this system (He spectrum plus laser parameters). The aim was to produce results where the control which can be effected by varying $\Delta t$ (see the interference peak corresponding to $E_{12}$), is demonstrable clearly.

As regards Fig. 6, where the field strengths are $I_1 = 1 \times 10^{14}$ W/cm² and $I_2 = 2 \times 10^{14}$ W/cm², now the ionization spectrum becomes rather complex, (broadened with additional peaks), in spite of the fact that the number of Rabi oscillations is, according to (10), only between 1 and 2. In this strong-field regime, the nonlinear effects and the interference among multiple transition paths involving discrete as well as scattering states, - see the discussion on the constitution of the $\Psi_{SSEA}(t)$ eq. 7 -, destroy the clear picture for control, exhibited by Figs. 2-5.

# IV.  CONCLUSION

Recent developments in experimental research on free-electron laser (FEL) show that it is now possible to produce simultaneously two ultrashort pulses with wavelengths in the XUV range,  e.g., [1, 5]. Although it appears that problems regarding their consistent characterization persist, (e.g., due to 'jitter'), there is realistic optimism about their systematic use in spectroscopic investigations of time-dependent phenomena associated with electron dynamics.

The theoretical work reported in this paper has proposed and shown computationally with reliable accuracy, that it is possible to use ultrashort *time delays*, $\Delta t$ , between two XUV pulses of ultrashort duration (about 15 to 80 cycles for wavelengths around 50-60 *nm*), and of moderate intensities, in order to *control* the dynamics of photoelectron emission based on a scheme that makes good use of interference of different excitation paths leading to the same final state.

Through analysis and trial calculations, we identified excitations in the Helium spectrum, (Fig. 1), and an appropriate range of pulse parameters, which demonstrate the argument quantitatively in the case of two- XUV photon, (58.4 *nm* and 52.22 *nm*), resonant ionization with few-femtosecond Gaussian pulses.

Evidently, if two ultrashort FEL pulses with shorter wavelengths are available for such synchronized interactions, the spectrum must be chosen accordingly. (E.g., larger energy differences in inner-shell resonant excitations or in excitations in the spectra of positive ions).

The predictions and interpretations are quantitative, and are based on the systematic solution of the fundamental equations 1, 2 nonperturbatively, according to the SSEA [12], as well as perturbatively, according to the results of lowest order perturbation theory, Eqs. (8, 9). The results are displayed in Figs. 2-6.

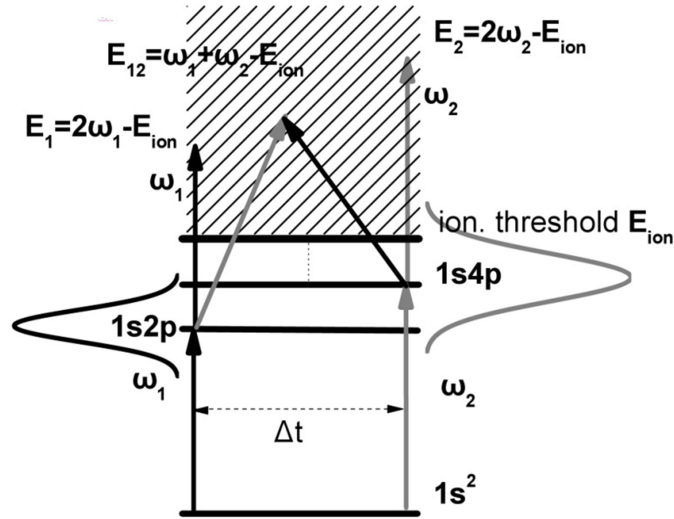

**Fig. 1**: The interference scheme in the two- XUV photon resonant ionization of Helium chosen for the present study of the possibility of using the ultrashort *time delay*, $\Delta t$, between the instants of application of the two ultrashort Gaussian XUV pulses, as a control knob. For the $1s2p\ ^1P^o$ state, $\omega_1$= 0.78 a.u. (58.4 *nm*), and for the $1s4p\ ^1P^o$ state, $\omega_2$= 0.87 a.u. (52.2 *nm*). The pulse durations are of the order of a few femtoseconds. The black and gray curves surrounding the two discrete levels represent realistically the energy profiles of the pulses, with central frequencies $\omega_1$ and $\omega_2$.

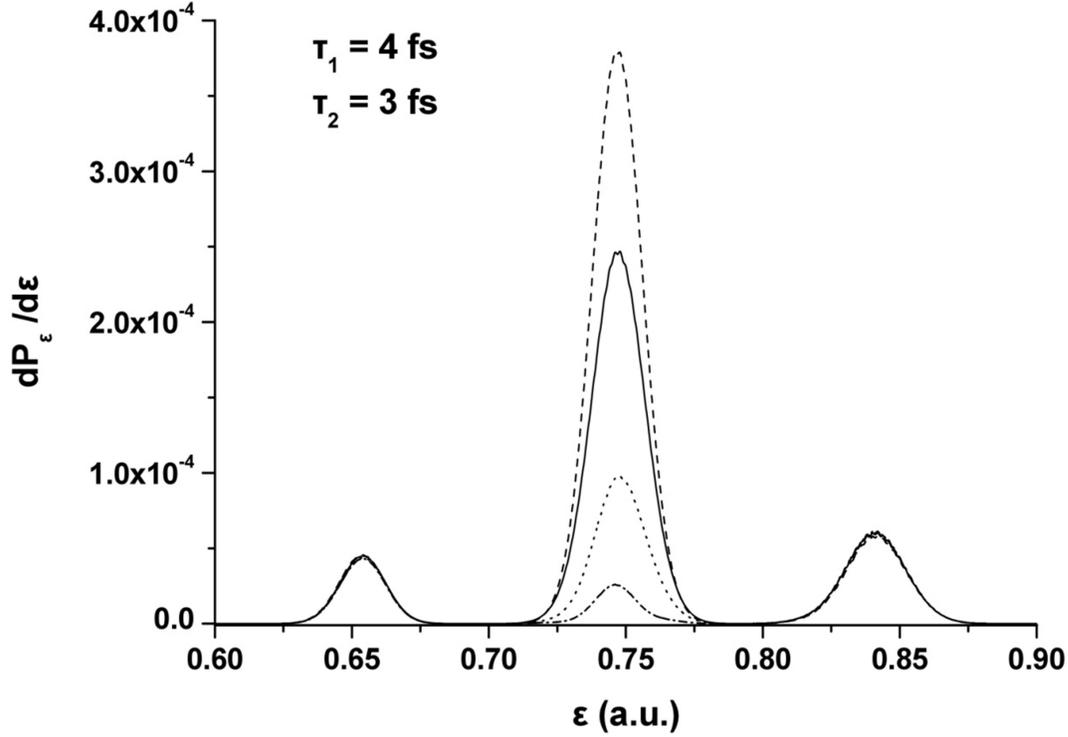

**Fig. 2**. Photoelectron spectrum for the two-XUV photon resonant ionization of Helium according to the scheme of Fig. 1. $\frac{dP_\varepsilon}{d\varepsilon}$ is the probability distribution per atomic unit of energy, defined and computed by Eq. (5). The full widths at half maximum, $\tau_1$ and $\tau_2$, are of the order of 20 cycles of each field. The curves correspond to positive and negative values of time delay, $\Delta t$. The solid line curve corresponds to $\Delta t = 0$ $fs$, the dashed one to $\Delta t = 2.4$ $fs$, the dotted one to $\Delta t = -2.4$ $fs$, and the dashed-dotted one to $\Delta t = -4.8$ $fs$. The left and right peaks represent the two-photon ionization of $He$ ionization, with $\omega_1 = 0.78$ a.u. (58.4 $nm$) and $\omega_2 = 0.87$ a.u. (52.2 $nm$). The middle peak is due to the interference at $E_{12}$. (See Fig. 1).

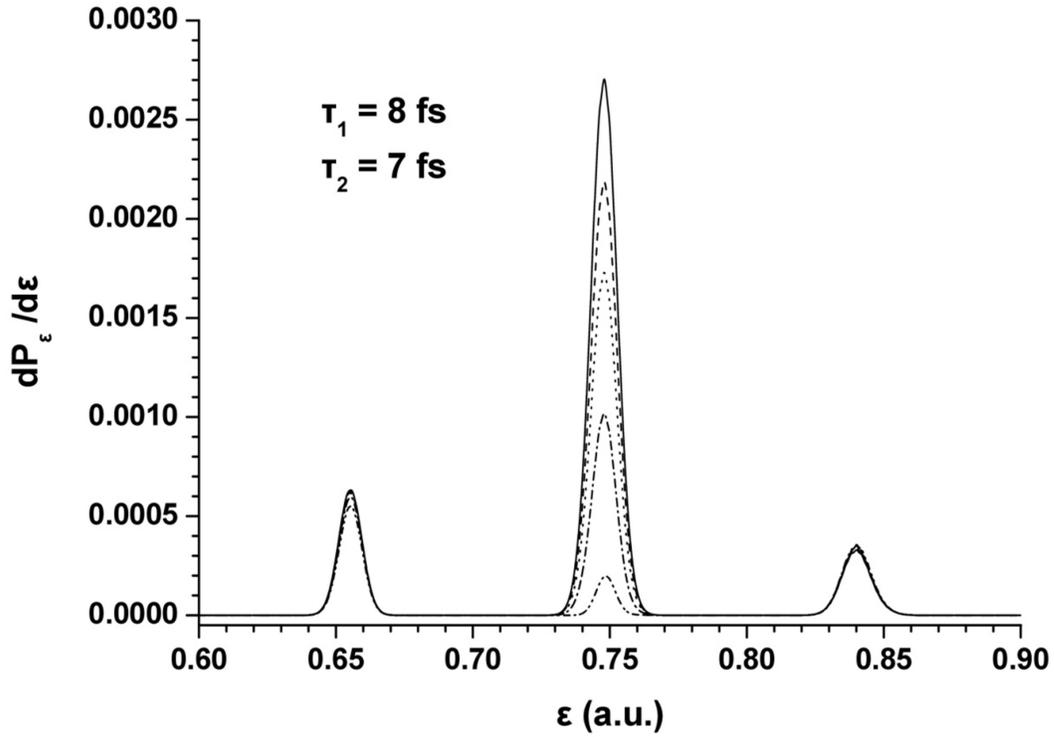

**Fig. 3**. As in Fig. 2, with $\tau_1$ and $\tau_2$ of the order of 40 cycles of each field. The curves are obtained for negative values of $\Delta t$. The solid line curve corresponds to $\Delta t = 0\ fs$, the dashed one to $\Delta t = -1.2\ fs$, the dotted one to $\Delta t = -2.4\ fs$, the dashed-dotted one to $\Delta t = -4.8\ fs$, and the dashed-dotted-dotted one to $\Delta t = -24.2\ fs$.

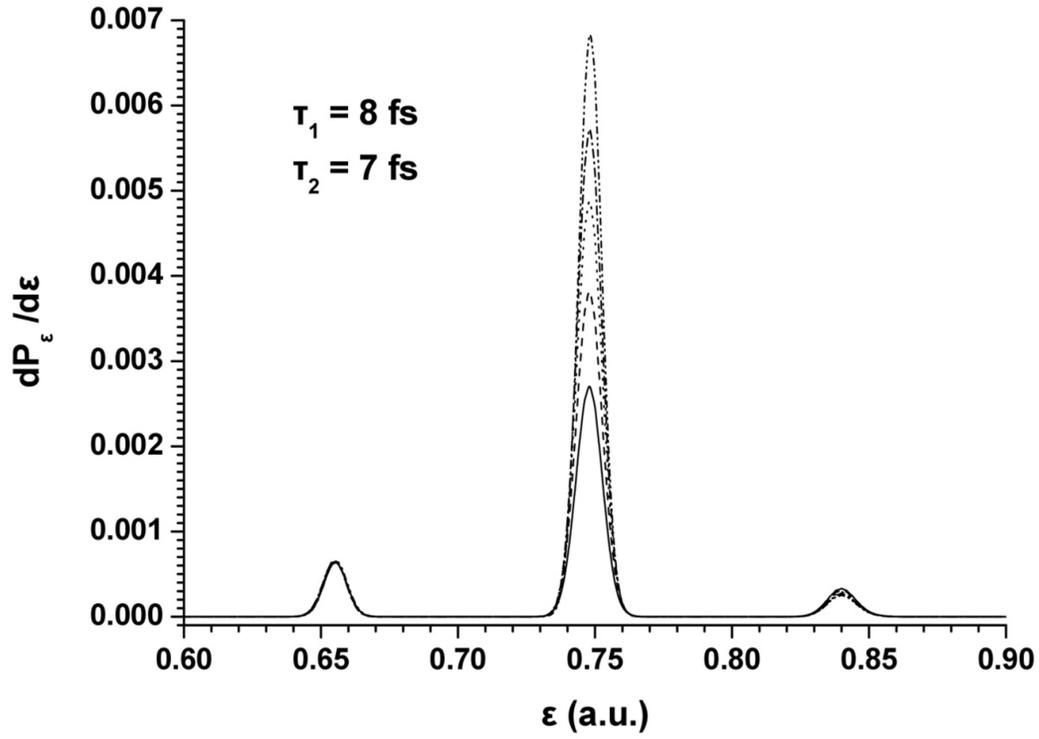

**Fig. 4**. As in Fig. 3, for positive values of $\Delta t$. The solid line curve corresponds to $\Delta t = 0 \; fs$, the dashed one to $\Delta t = 2.4 \; fs$, the dotted one to $\Delta t = 4.8 \; fs$, the dashed-dotted one to $\Delta t = 7.3 \; fs$, and the dashed-dotted-dotted one to $\Delta t = 16.9 \; fs$.

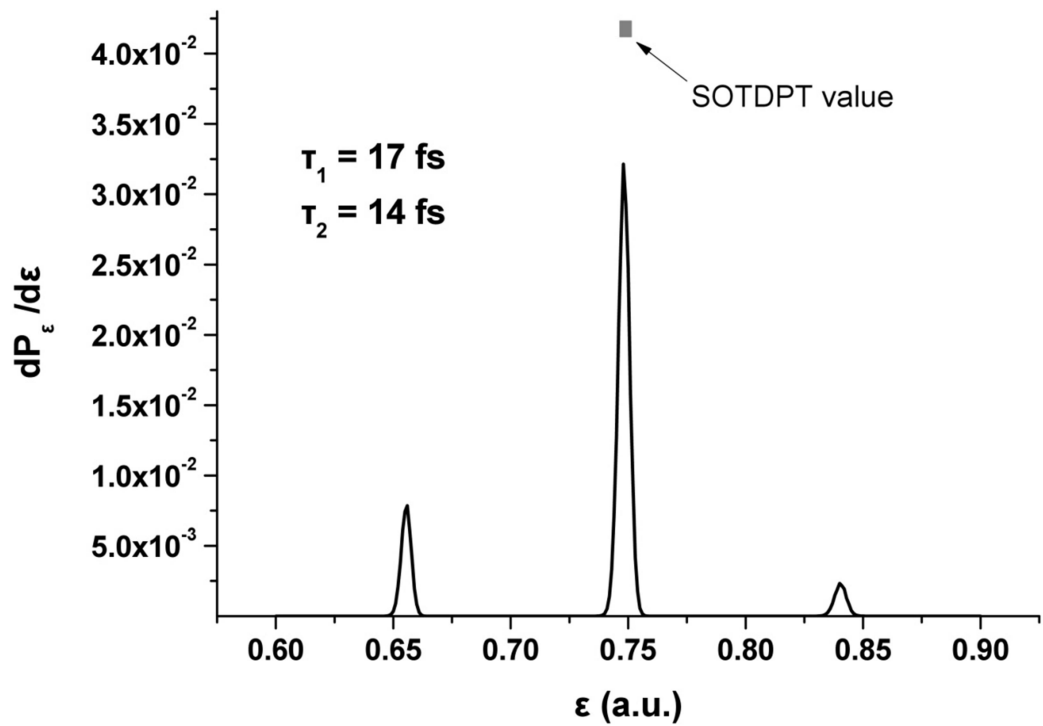

**Fig. 5**. As in Fig. 2, with $\tau_1$ and $\tau_2$ of the order of 80 cycles of each field. The curves correspond to $\Delta t = 0$. The gray square shows the value calculated from the second order time-dependent perturbation theory (SOTDPT). See text.

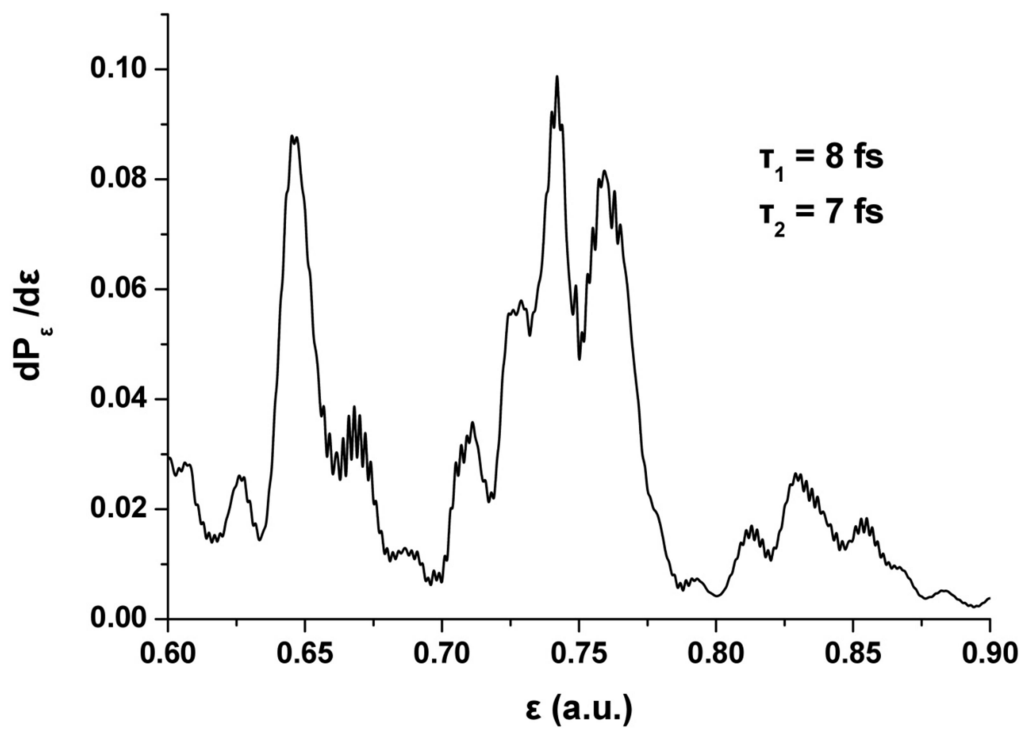

**Fig. 6**: As in Fig. 3, for intensities $I_1 = 1 \times 10^{14}\,W/cm^2$ and $I_2 = 2 \times 10^{14}\,W/cm^2$ . The curve corresponds to $\Delta t = 0$. The SOTDPT value for the interference (middle) peak is 6.6, i.e., an order of magnitude off the value obtained by the SSEA nonperturbative calculation.